\documentclass[%
 reprint,
superscriptaddress, aps,
preprintnumbers,
nofootinbib,
]{revtex4-2}

\usepackage{graphicx}
\usepackage{bm}
\usepackage{amssymb}
\usepackage{lipsum}
\usepackage{amsmath}
\usepackage[colorlinks,citecolor=blue,urlcolor=blue,linkcolor=blue]{hyperref}
\usepackage{hyphenat}
\usepackage{microtype}
\usepackage{mathtools}
\usepackage[utf8]{inputenc}
\usepackage[T1]{fontenc}
\usepackage{xcolor}
\usepackage{arydshln} 
\usepackage{lineno}
\usepackage{orcidlink}
\usepackage{comment}

\newcommand{\MGvATNLO}{{\tt {\sc MadGraph5}\_aMC@NLO}}

\newcommand{\ttz}{$t \bar{t} Z$ }

\begin{document}

\preprint{ICPP-96}

\title{Analyzing $t\bar{t}Z$-couplings at the future $e^-p$ collider}


\author{Katlego Machethe\,\orcidlink{0000-0001-6980-0952}}
\email{kmi.machethe@gmail.com}
\affiliation{School of Physics and Institute for Collider Particle Physics, University of the Witwatersrand, Johannesburg, Wits 2050, South Africa.}
\affiliation{iThemba LABS, National Research Foundation, Somerset West, South Africa.}

\author{Pramod Sharma\,\orcidlink{0000-0001-6381-7876}}
\email{pramodsharma.iiser@gmail.com}
\affiliation{Indian Institute of Science Education and Research, Knowledge City, Sector 81, S. A. S. Nagar, Manauli PO 140306, Punjab, India.}
\affiliation{School of Physics and Institute for Collider Particle Physics, University of the Witwatersrand, Johannesburg, Wits 2050, South Africa.}

\author{Mukesh Kumar\,\orcidlink{0000-0003-3681-1588}}
\email{mukesh.kumar@cern.ch}
\affiliation{School of Physics and Institute for Collider Particle Physics, University of the Witwatersrand, Johannesburg, Wits 2050, South Africa.}

\author{Rafiqul Rahaman\,\orcidlink{0000-0002-3907-829X}}
\email{rafiqul@if.usp.br}
\affiliation{Instituto de Física, Universidade de São Paulo, Rua do Matão 1371, Butanta, São Paulo, 05508-220, SP,
Brazil.}

\author{Bruce Mellado\,
\orcidlink{0000-0003-4838-1546}}
\email{bmellado@cern.ch}
\affiliation{School of Physics and Institute for Collider Particle Physics, University of the Witwatersrand, Johannesburg, Wits 2050, South Africa.}
\affiliation{iThemba LABS, National Research Foundation, Somerset West, South Africa.}


\begin{abstract}
The proposed Large Hadron electron Collider (LHeC), with a center-of-mass energy of $\sqrt{s} \approx 1.3$~TeV, provides a clean and sensitive environment to probe the top quark's neutral current interactions with the $Z$ boson via $e^- p \to e^- t \bar{t}$. We investigate the precision with which the Standard Model (SM) $t\bar{t}Z$ couplings $-$ the vector and axial-vector components ($\Delta C_{1V},\, \Delta C_{1A}$) can be measured, along with possible new physics effects parametrized by higher-dimensional operators inducing weak electric and magnetic dipole-like interactions ($C_{2V},\, C_{2A}$). Focusing on the semileptonic decay channel, where either the top quark decays leptonically to a positively charged lepton ($\ell^+ = e^+, \mu^+$) and the anti-top hadronically, or vice versa, we utilize the azimuthal angle difference $\Delta \phi_{e^- \ell^\pm}$ between the scattered electron and the charged lepton $\ell^\pm$ as the key observable to derive projected constraints. Using a one-parameter multi-bin $\chi^{2}$ analysis of the differential $\Delta \phi_{e^- \ell^\pm}$ distribution, the constraints on $\Delta C_{1A}$ improve from $\mathcal{O}(10^{-1})$ at $50~\mathrm{fb}^{-1}$ to $\mathcal{O}(10^{-2})$ at $1000~\mathrm{fb}^{-1}$, while the improvement for $\Delta C_{1V}$ remains within the same order of magnitude, $\mathcal{O}(10^{-1})$. These results correspond to average relative precisions of approximately 8\% and 68\%, respectively, with respect to their SM values. The anomalous tensor couplings $C_{2V}$ and $C_{2A}$ are constrained at the $\mathcal{O}(10^{-1})$ level even at low luminosity and improve moderately with increasing luminosity. While the two-parameter analysis broadens the allowed regions due to parameter correlations, it maintains competitive sensitivity, particularly for the SM-like couplings. Throughout, we assume a systematic uncertainty of $\delta_s = 5\%$ in all analyses, and all results are quoted at the 95\% confidence level. These results demonstrate the LHeC’s potential to provide complementary and competitive sensitivity to top–$Z$ couplings compared to current and future hadron and lepton collider capabilities.
\end{abstract}

\maketitle

\section{Introduction}

The top quark, being the heaviest particle in the Standard Model (SM), plays a crucial role in probing the mechanism of electroweak symmetry breaking (EWSB) and in searching for physics beyond the SM (BSM). Precise measurements of top quark couplings, particularly its neutral current interactions with the $Z$ boson, offer a sensitive window into new physics, as such interactions can be significantly modified by effects from higher-dimensional operators or heavy new states~\cite{Aguilar-Saavedra:2008nuh,Aguilar-Saavedra:2012ehg,Rontsch:2014cca,Rontsch:2015una,ATLAS:2019fwo,CMS:2019too,CMS:2022hjj,ATLAS:2023eld}.

The production of top-quark pairs in association with a $Z$ boson ($t\bar{t}Z$) has been actively studied at the Large Hadron Collider (LHC), both theoretically and experimentally. The ATLAS and CMS Collaborations have reported increasingly precise measurements of the $t\bar{t}Z$ production cross section at $\sqrt{s} = 13$~TeV in multi-lepton final states~\cite{ATLAS:2019fwo, ATLAS:2023eld,CMS:2019too, CMS:2021ugv,CMS:2024mke,CMS:2022hjj, CMS:2024mke}, which are consistent with SM predictions~\cite{Frederix:2011zi, Lazopoulos:2008de, Campbell:2013yla}, yet still allow room for moderate deviations that could arise from new physics. Effective field theory (EFT) approaches have been employed to constrain anomalous $t\bar{t}Z$ couplings~\cite{Buckley:2015lku, Durieux:2018tev, Cao:2020npb, Cao:2015qta}, including those arising from weak electric and magnetic dipole moment operators~\cite{Aguilar-Saavedra:2008nuh,Zhang:2010dr}. Next-to-leading order (NLO) QCD calculations provide precise predictions for inclusive and differential cross sections, which are essential for interpreting experimental measurements~\cite{Rontsch:2014cca}. 

Future electron–proton colliders such as the proposed Large Hadron electron Collider (LHeC) offer a promising avenue to complement hadron and lepton collider programs~\cite{LHeCStudyGroup:2012zhm,Ahmadova:2025vzd,Klein:2024mao}. With a center-of-mass energy of $\sqrt{s} \approx 1.3$~TeV and clean experimental environment, the LHeC is expected to provide unique sensitivity to electroweak couplings~\cite{Dutta:2013mva,Behera:2018ryv, Britzger:2020kgg}. Unlike the LHC, where the QCD‑dominated environment suffers from high pile‑up and complex hadronic backgrounds that obscure event reconstruction, the $e^-p$ collisions at the LHeC occur in an ultra‑clean setting with negligible pile‑up, low hadronic noise, and well‑controlled initial‑state kinematics $-$ thus enabling far more precise reconstruction of observables~\cite{Biswal:2012mp,Kumar:2015kca,Kumar:2015tua, Andre:2022xeh,Mosomane:2017jcg,Coleppa:2017rgb}, and significantly enhanced sensitivity to subtle deviations in the $t\bar{t}Z$ couplings through neutral current processes such as $e^- p \to e^- t \bar{t}$. 

In the SM, the interaction of the top quark with the $Z$ boson is governed by a vector and axial-vector structure, reflecting the chiral nature of the electroweak interaction. The relevant interaction term in the SM Lagrangian is given by:
\begin{equation}
\mathcal{L}_{t\bar{t}Z}^{\mathrm{SM}} \supset \frac{g}{2\cos\theta_W} \, \bar{t} \gamma^\mu \left( C_{1V}^{\mathrm{SM}} - C_{1A}^{\mathrm{SM}} \gamma^5 \right) t \, Z_\mu, \label{sm-ttz}
\end{equation}
where $C_{1V}^{\mathrm{SM}} = \frac{1}{2} - \frac{4}{3} \sin^2\theta_W \approx 0.192$ and $C_{1A}^{\mathrm{SM}} = \frac{1}{2}$ denote the vector and axial-vector couplings of the top quark to the $Z$ boson, with the weak mixing angle $\theta_{W}$ corresponding to $\sin^2\theta_W \approx 0.231$.

In this work, we aim to study the sensitivity of the $t\bar{t}Z$ vertex to possible deviations from this structure. Such deviations can arise from higher-dimensional operators that encode the effects of new physics at higher energy scales. In particular, the $t\bar{t}Z$ vertex is sensitive to anomalous weak dipole moments of the top quark, which induce tensorial interactions beyond the SM. These corrections are typically captured in an effective field theory framework through terms of the form $\bar{t} \sigma^{\mu\nu} q_\nu Z_\mu$, corresponding to magnetic and electric dipole operators. Precise measurements of these couplings therefore provide a direct probe of new physics in the top-quark electroweak sector, complementary to direct searches at high-energy colliders.

The structure of this article is as follows: In \autoref{EFT-theory}, we present the theoretical formalism based on the interaction Lagrangian used to study the sensitivity of the $t\bar{t}Z$ couplings within both the Standard Model and its extensions. The corresponding studies on signal and background processes in the $e^-p$ collider environment are detailed in \autoref{ttZ_production}. In \autoref{analysis}, we present the analysis methodology and the results of our study. Finally, we summarize and conclude our findings in \autoref{sec:Conc}.

\section{Formalism}
\label{EFT-theory}

To study potential deviations from the SM in the $t\bar{t}Z$ vertex, we consider a general Lorentz-invariant parameterization that includes both the SM-like vector and axial-vector couplings as well as anomalous dipole interactions arising from dimension-six operators, $C_i$ with $i \in \{1V, 1A, 2V, 2A\}$. The effective interaction Lagrangian is written as~\cite{Grzadkowski:2010es,Aguilar-Saavedra:2008nuh,Aguilar-Saavedra:2012ehg, Rahaman:2022dwp}:
\begin{align}
\mathcal{L}_{t\bar{t}Z} \supset \frac{g}{2\cos\theta_W} \, &\bar{t} \Big[
     \gamma^\mu \left( \Delta C_{1V} - \Delta C_{1A} \gamma^5 \right) \notag \\
    & + \frac{i \sigma^{\mu\nu} q_\nu}{m_Z} \left( C_{2V} - i C_{2A} \gamma^5 \right)
\Big] t \, Z_\mu.
\label{eq:ttZ-Lag}
\end{align}
In this expression, $\Delta C_{1V}$ and $\Delta C_{1A}$ include possible deviations from the SM vector and axial-vector couplings, while $C_{2V}$ {(CP-even) and $C_{2A}$ (CP-odd) represent anomalous weak magnetic and electric dipole couplings, which are zero at tree level in the SM and typically arise from higher-dimensional operators. Here, $q_\nu$ denotes the four-momentum transfer carried by the $Z$ boson, $\sigma^{\mu\nu} = \frac{i}{2}[\gamma^\mu, \gamma^\nu]$, and the dipole terms are normalized by the electroweak scale $\Lambda = m_Z$. Hence, the working Lagrangian for this study is taken as ${\cal L}_{\rm tot} = {\cal L}_{\rm SM} + {\cal L}_{t\bar{t}Z}$, where ${\cal L}_{\rm SM}$ denotes the full SM Lagrangian. This Lagrangian is implemented in the {\tt FeynRules} package~\cite{Alloul:2013bka} to construct the necessary model files.

In the context of effective field theory (EFT), the effects of new physics arising from higher-dimensional operators $\mathcal{O}_{ij}$ can be modeled through the effective Lagrangian:
\begin{equation}
    \mathcal{L}_{\text{eff}} = \mathcal{L}_{\text{SM}} + \sum_{i,n} \frac{C_{ij}}{\Lambda^{n-4}} \mathcal{O}_{ij},
\end{equation}
where $n > 4$, $C_{ij}$ are the corresponding Wilson coefficients (WCs), and $\Lambda$ denotes the energy scale up to which the EFT description remains valid. In the Warsaw basis, the higher-dimensional operators relevant to the $t\bar{t}Z$ interaction are~\cite{Grzadkowski:2010es,BessidskaiaBylund:2016jvp}:
\begin{equation}
\left.
\begin{aligned}
    \mathcal{O}_{tB} &= \big( \bar{Q} \sigma^{\mu \nu} t \big) \bar{\phi} B_{\mu \nu},\\
    \mathcal{O}_{tW} &= \big( \bar{Q} \sigma^{\mu \nu} \tau^{I} t \big) \bar{\phi} W_{\mu \nu}^{I},\\
    \mathcal{O}_{\phi t} &= \big( \phi^{\dagger} i \overleftrightarrow{D_{\mu}} \phi \big) \big( \bar{t} \gamma^{\mu} t \big),\\
    \mathcal{O}_{\phi Q}^{(1)} &= \big( \phi^{\dagger} i \overleftrightarrow{D_{\mu}} \phi \big) \big( \bar{Q} \gamma^{\mu} Q \big),\\
    \mathcal{O}_{\phi Q}^{(3)} &= \big(\phi^{\dagger} i \overleftrightarrow{D_{\mu}} \tau^{I} \phi \big) \big(\bar{Q} \gamma^{\mu} \tau^{I} Q \big)
\end{aligned}
\right\}
\label{operators}
\end{equation}
where $Q$ and $t$ denote the left-handed top-bottom quark doublet and the right-handed top-quark singlet, respectively. The $\tau^{I}$ are the Pauli matrices, and $\phi$ is the Higgs doublet, with $\tilde{\phi} \equiv i \tau^2 \phi$ its charge-conjugated counterpart. The couplings in Lagrangian~\eqref{eq:ttZ-Lag} can be expressed in terms of the effective parameters $c_{tZ}$, $c_{tZ}^{I}$, $c_{\phi t}$, and $c^{-}_{\phi Q}$:
\begin{equation}
\left.
\begin{aligned}
    &\Delta C_{1V} =  \frac{v^{2}}{ \Lambda^{2} } \mathfrak{Re} \big[ -c_{\phi t} - c^{-}_{\phi Q}  \big], \\
    &\Delta C_{1A} =  \frac{v^{2}}{ \Lambda^{2}} \mathfrak{Re} \big[ -c_{\phi t} + c^{-}_{\phi Q}  \big], \\
    &C_{2V} = \frac{\sqrt{2}\, v^{2}}{ \Lambda^{2} } c_{tZ}, \quad
    C_{2A} = \frac{\sqrt{2}\, v^{2}}{ \Lambda^{2} } c_{tZ}^{I},
\end{aligned}
\right\}
\label{eq:cwc_all}
\end{equation}
where $v \simeq 246$~GeV is the Higgs vacuum expectation value. The effective parameters above are, in turn, related to the WCs $C_{ij}$ of the higher-dimensional operators $\mathcal{O}_{ij}$ and given as~\cite{Aguilar-Saavedra:2018ksv}:
\begin{align}
    c_{tZ} =&\, \mathfrak{Re} (- \sin\theta_W\cdot C_{tB} + \cos\theta_W \cdot C_{tW} ),\label{wc1}\\
    c_{tZ}^{I} =&\, \mathfrak{Im} (- \sin\theta_W \cdot C_{tB} + \cos\theta_W\cdot C_{tW} ),\label{wc2}\\
    c_{\phi t} =&\, C_{\phi Q},\label{wc3} \\
    c^{-}_{\phi Q} =&\, C^{(1)}_{\phi Q} - C^{(3)}_{\phi Q}.
    \label{wc4}
\end{align}
From eq.\eqref{wc1} and eq.\eqref{wc2}, it is evident that the weak magnetic and electric dipole moments of the top quark arise from the real and imaginary parts of the Wilson coefficient combination ($C_{tB}$, $C_{tW}$) respectively. On the other hand, $c_{\phi t}$ and $c_{\phi Q}$ contribute to modifications of the top quark's neutral current interactions with the $Z$ boson.

The 95\% confidence level (C.L.) bounds on the relevant WCs, assuming $\Lambda = 1~\mathrm{TeV}$, have been derived by the CMS and ATLAS collaborations at $\sqrt{s} = 13~\mathrm{TeV}$. These constraints are obtained from EFT interpretations of top quark pair production with a $Z$ boson ($t\bar{t}Z$) using various channels and observables. CMS has reported constraints based on two datasets with different integrated luminosities and analysis channels. For an integrated luminosity of $\mathcal{L} = 77.5~\mathrm{fb}^{-1}$ in the $3\ell$ channel~\cite{CMS:2019too}, using the observables $p^Z_T$ (transverse momentum of the $Z$ boson) and $\cos \theta^*_Z$ (cosine of the angle between the negatively charged lepton and the $Z$ boson direction in the $Z$ boson rest frame), the 95\% C.L. bounds are:
\begin{align}
c_{\phi t} / \Lambda^2 &\in [0.3, 5.4], \quad c^-_{\phi Q} / \Lambda^2 \in [-4.0, 0.0], \notag \\
c_{tZ} / \Lambda^2 &\in [-1.1, 1.1], \quad c^I_{tZ} / \Lambda^2 \in [-1.2, 1.2]. \notag
\end{align}
For $\mathcal{L} = 138~\mathrm{fb}^{-1}$ in the $t\bar{t}Z$, $Z \to b \bar{b}$ channel~\cite{CMS:2022hjj}, the updated 95\% C.L. bounds are:
\begin{align}
c_{\phi t} / \Lambda^2 &\in [-12, 6.3], \quad c^-_{\phi Q} / \Lambda^2 \in [-6.6, 8.7], \notag \\
c_{tZ} / \Lambda^2 &\in [-1.0, 1.1]. \notag
\end{align}
ATLAS has provided constraints based on an $\mathcal{L} = 140~\mathrm{fb}^{-1}$~\cite{ATLAS:2023eld}. These bounds are derived using an EFT interpretation with Bayesian statistics, employing a multimodal Gaussian likelihood function. The analysis uses fiducial $t\bar{t}Z$ cross sections and normalized differential distributions at the parton level in the combined $3\ell$ and $4\ell$ final states. The observables include the transverse momentum of the $Z$ boson ($p^Z_T$), the absolute rapidity of the $Z$ boson ($|y^Z|$), $\cos \theta^*_Z$ (defined as above), the transverse momentum of the top quark ($p^t_T$), the azimuthal separation between the $t\bar{t}$ system and the $Z$ boson ($|\Delta \phi_{t\bar{t}, Z}|$), and the absolute rapidity of the $t\bar{t}Z$ system ($|y^{t\bar{t}Z}|$). The 95\% C.L. bounds, obtained from quadratic fits where each WC is constrained independently while fixing others to zero, are:
\begin{align}
c_{\phi t} / \Lambda^2 &\in [-2.2, 1.6], \quad c^-_{\phi Q} / \Lambda^2 \in [-1.16, -0.45], \notag \\
c_{tZ} / \Lambda^2 &\in [0.06, 0.13], \quad c^I_{tZ} / \Lambda^2 \in [-0.01, 0.01]. \notag
\end{align}

These experimental bounds serve as benchmarks for evaluating the sensitivity of the LHeC in probing $t\bar{t}Z$ couplings through signal and background analyses based on the sensitive observables discussed in the following sections.

\section{Signal and Background Simulation}
\label{ttZ_production}

To probe the sensitivities of the \ttz\ couplings as in the Lagrangian~\eqref{eq:ttZ-Lag} within the $e^-p$ environment, we simulate signal events through the neutral-current ({\tt NC}) process $e^{-} p \to e^{-} t \bar{t}$, with representative Feynman diagrams shown in \autoref{fig:feyn}. In the SM, this process proceeds via a $t$-channel exchange of $\gamma$ and $Z$ bosons emitted from the incoming electron line, as illustrated in the leading-order diagrams in \autoref{fig:feyn}~(a). The contribution from new physics affecting the \ttz\ coupling, as introduced in Lagrangian~\eqref{eq:ttZ-Lag}, is shown in \autoref{fig:feyn}~(b). All simulations are performed at parton level using electron and proton beam energies of $E_e = 60$~GeV and $E_p = 7$~TeV, respectively, corresponding to a center-of-mass energy of $\sqrt{s} \approx 1.3$~TeV, as proposed for the LHeC~\cite{LHeCStudyGroup:2012zhm}, using the \MGvATNLO\ package~\cite{Alwall:2011uj}. We consider an electron beam  polarized at $-80\%$.

\begin{figure}[t]
    \centering
    \begin{minipage}[b]{0.22\textwidth}
        \includegraphics[width=\textwidth]{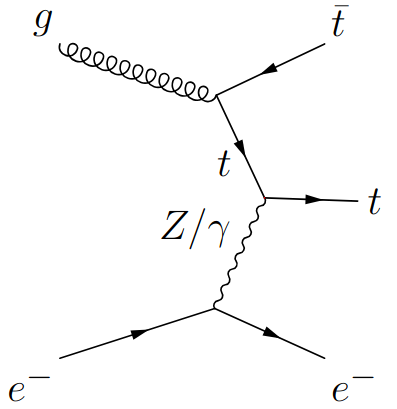}
        \centerline{(a)}
    \end{minipage}
    \hfill
    \begin{minipage}[b]{0.22\textwidth}
        \includegraphics[width=\textwidth]{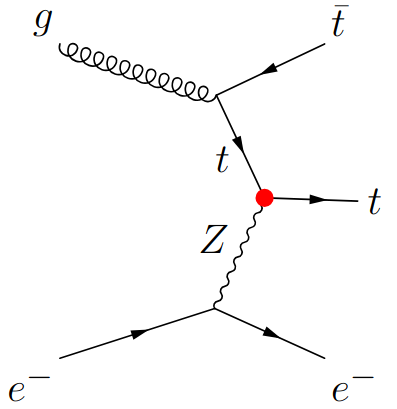}
        \centerline{(b)}
    \end{minipage}
    \caption{Representative leading-order Feynman diagrams for the process $e^- p \to e^- t\bar{t}$: 
    (a) the SM contributions mediated by $\gamma$ and $Z$ bosons, and 
    (b) the new physics contribution probing the $t\bar{t}Z$ coupling, 
    where the $t\bar{t}Z$ vertex is defined in Lagrangian~\eqref{eq:ttZ-Lag}.}
    \label{fig:feyn}
\end{figure}

\begin{table}
\centering
\caption{Cross sections $\sigma$ in fb for the process $e^{-} p \to e^{-} t \bar{t}$ in the semileptonic, hadronic, and dileptonic final states. The first row shows the SM prediction. The remaining rows correspond to new physics scenarios where one $t\bar{t}Z$ coupling is set to $= 0.5$ while the other three couplings are fixed to zero. Note that SM–new physics interference is included in all cross section estimates.}
\begin{ruledtabular}
\begin{tabular}{rccc}
        Coupling &  Semi-leptonic & Hadronic & Di-leptonic \\ \hline 
        SM signal & 4.39 & 6.60 & 0.73 \\
        $\Delta C_{1V} = 0.5$ & 5.54 & 8.33 & 0.92 \\
        $\Delta C_{1A} = 0.5$ & 9.34 & 14.1 & 1.56 \\
        $C_{2V} = 0.5$ & 7.06 & 10.6 & 1.18 \\
        $C_{2A} = 0.5$ & 7.16 & 10.8 & 1.19 
\end{tabular}
\end{ruledtabular}
\label{tab:sm_bsm_xs}
\end{table}

\begin{figure}[t]
     \centering
    \includegraphics[width=1\linewidth]{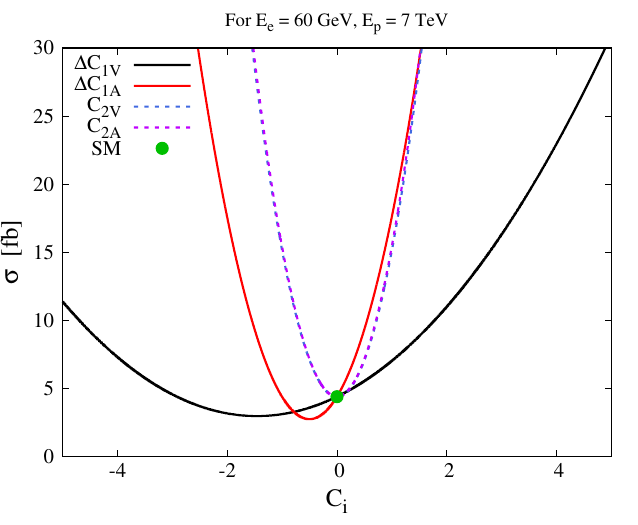} 
     \caption{Total cross section for the process $e^{-} p \to e^{-} t \bar{t}$ in the semi-leptonic final state, shown as a function of each $t\bar{t}Z$ coupling parameter $C_i$, with all other couplings set to zero. The results are obtained for $\sqrt{s} \approx 1.3~\rm{TeV}$ at the LHeC.}
     \label{inclusive_xsecs}
 \end{figure}

\begin{table*}[t]
    \caption{Cutflow for the SM signal and dominant backgrounds normalized to the luminosity ${\mathcal{L}} = 50~ (1000)~\rm fb^{-1}$, and assuming a $b$-tagging efficiency of 70\%. The final column corresponds to the optimized selection cuts: $-5 < \eta_{e^{-}} < 1$, $\eta_{b} > -3.5$, $60 < m_{jj} < 100~\rm GeV$, and $m_{bjj} < 90~\rm GeV$. The local significance $N_{\rm SD}$  with a systematic uncertainty of $\delta_s = 5\%$ for the SM signal and backgrounds is shown in the last row.}
    \label{semilepton_cutflow}
\begin{ruledtabular}
\begin{tabular}{llll}
Process & Generation level & $p_T^b > 20~\rm GeV$ & Optimized cuts \\
\hline
SM signal & 108 (2153) & 98 (1967) & 79 (1587) \\
\hline
$e^{-} p \to e^{-} t W b$ (No top line) & 152 (3049) & 44 (885) & 17 (340) \\
\hline
$e^{-} p \to e^{-} W^{+} b \bar{b} j j$ (No top/$W^{-}$ line) & 149 (2972) & 37 (737) & 0.1 (1.9) \\
\hline
$e^{-} p \to e^{-} W^{+} W^{-} b \bar{b}$ (No top line) & 1.1 (21) & 0.2 (5) & 0.1 (3) \\
\hline
$N_{\rm SD}$ ($\delta_s = 5\%$) & 4.2 (6.5) & 6.6 (14) & 7.5 (17.2) \\
\end{tabular}
\end{ruledtabular}
\end{table*}

With the top (anti-top) quark decaying exclusively through the $W^{+}b$ ($W^{-}\bar{b}$) channel, with a branching ratio of ${\tt BR}(t \to W^{+} b) \equiv {\tt BR}(\bar{t} \to W^{-} \bar{b}) \simeq 100\%$, the final state topology of $t\bar{t}$ events is determined by the decay modes of the $W^\pm$ bosons. This gives rise to three distinct final states: \textit{fully hadronic} ($W^\pm\to jj$), \textit{semi-leptonic} ($W^{+}\to \ell^{+}\bar{\nu}_{\ell}/jj$ and $W^{-}\to jj/\ell^{-}\nu_{\ell}$),  and \textit{di-leptonic} ($W^\pm\to \ell^\pm\nu_{\ell}$); where $\ell = e,\, \mu$.  A comparative study of the SM and representative BSM couplings ($=0.5$) across all three final states is summarized in \autoref{tab:sm_bsm_xs}. For each new physics benchmark, only one coupling is set to 0.5, while the others are fixed to zero. The total cross section includes contributions from both the SM and its interference with the new physics. As observed, the cross section for the semi-leptonic channel is sub-dominant, while the fully hadronic channel is dominant. This is primarily due to the larger \texttt{BR} of the $W$ boson decaying hadronically ($W\to jj$), which is approximately 67\%, compared to leptonic decays ($W\to \ell \nu_\ell$), which occur with a \texttt{BR} of about 11\% per lepton flavor. The di-leptonic channel yields the lowest cross section, differing by roughly an order of magnitude, since it requires both $W$ bosons to decay leptonically.

For all event samples, the following generator-level cuts were applied to ensure well-separated and detectable final-state objects: transverse momentum $p_T^{j} > 20~{\rm GeV}$ for light jets and $p_T^{\ell} > 10~{\rm GeV}$ for charged leptons; pseudorapidity $|\eta_{j}| < 5$ for jets and $-5 < \eta_{\ell} < 2.5$ for leptons. Additionally, a minimum angular separation of $\Delta R > 0.4$ was required between any pair of final-state objects (jets and/or leptons), where $\Delta R = \sqrt{(\Delta \eta)^2 + (\Delta \phi)^2}$ combines the differences in pseudorapidity and azimuthal angle.

Among the three channels, the semi-leptonic final state was chosen for our detailed analysis. This channel provides a balanced compromise: it benefits from a moderate branching ratio, offers cleaner reconstruction compared to the fully hadronic channel (which suffers from large QCD multijet backgrounds), and retains more statistics than the dileptonic channel (which is suppressed by the leptonic branching ratio of both $W$ bosons). Moreover, the presence of exactly one charged lepton in the final state significantly improves the event selection efficiency and helps in suppressing reducible backgrounds, making it optimal for probing the $t\bar{t}Z$ couplings in the $e^{-}p$ environment.

The signal process is given by: $e^{-} p \to e^{-} t \bar{t}$, where $t \to b\, (W^{+} \to \ell^{+} \nu_{\ell}/jj)$ and $\bar{t} \to \bar{b}\, (W^{-} \to j j/\ell^{-} \bar{\nu}_{\ell})$. In \autoref{inclusive_xsecs}, we show the total cross section for this process as a function of each individual \ttz coupling parameter $C_i$, varied one at a time while keeping all others fixed to zero. The cross section exhibits a characteristic parabolic dependence on each coupling, resulting from the interplay between the linear interference terms (from SM--BSM interference) and the purely quadratic BSM contributions. Notably, for values $-2.917 < \Delta C_{1V} < 0$ and $-1.001 < \Delta C_{1A} < 0$, the cross section decreases further below the SM prediction. This reduction arises from destructive interference between the SM amplitude and the contributions from modified coupling values. As the deviations grow larger, the quadratic terms become increasingly significant, leading to a reversal in the trend and an eventual increase in the total cross section. Since $C_{2V}$ and $C_{2A}$ arise from momentum-dependent, higher-dimensional operators, smaller values of these couplings are sufficient to produce the same cross section as larger values of the SM-like couplings $\Delta C_{1V}$ and $\Delta C_{1A}$.

The dominant backgrounds contributing to this final state include:
\begin{itemize}
    \item[(a)] $e^{-} p \to e^{-} t W^{-} b$ or $e^{-} \bar{t} W^{+} \bar{b}$ (single top production),
    \item[(b)] $e^{-} p \to e^{-} W^{+} b \bar{b} j j$ (non-resonant $W$ production),
    \item[(c)] $e^{-} p \to e^{-} W^{+} W^{-} b \bar{b}$ (diboson with heavy flavor),
    \item[(d)] and other reducible backgrounds involving jets misidentified as leptons or $b$-jets.
\end{itemize}
We note that, after applying miss-tagging probabilities of $\epsilon_c = 0.1$ for charm jets and $\epsilon_j = 0.01$ for light-flavor jets being misidentified as $b$-jets, the contribution of the corresponding background processes becomes negligible. Furthermore, a $b$-tagging efficiency of 70\% is assumed uniformly for both the signal and the dominant background channels.

To enhance the signal-to-background ratio, we apply a set of optimized selection cuts targeting the kinematic features of the final-state $b$-jets and invariant mass distributions. Specifically, we require the transverse momentum of $b$-jets to satisfy $p_T^{b} > 20~\rm GeV$, with their pseudorapidity constrained by $\eta_{b} > -3.5$, while the scattered electron is required to lie within $-5 < \eta_{e^-} < 1$. To distinguish the scattered $e^-$ from the $e^-$ originating from top-quark decay, electrons are sorted by $\eta$, as the scattered electron typically has the highest $\eta$. For the hadronically decaying $W$ boson, we impose an invariant mass window of $60~{\rm GeV} < m_{jj} < 100~{\rm GeV}$. To reconstruct the hadronically decaying top and anti-top quarks, we further demand $m_{bjj} > 90~{\rm GeV}$.


To evaluate the effectiveness of the selection strategy, we compute the local statistical significance, denoted as $N_{\rm SD}$, using the signal ($S$) and background ($B$) event yields at a given integrated luminosity ${\cal L}$. The significance is calculated according to the expression~\cite{Mosala:2023sse}:
\begin{equation}
\label{eq:significance}
N_{\rm SD} = \frac{S}{\sqrt{S + B + (\delta_{s}\cdot S)^2 + (\delta_{s}\cdot B)^2}}.
\end{equation}
Here, $\delta_{s}$ represents the combined systematic and statistical uncertainty, which we take to be 5\% for our benchmark integrated luminosity of ${\cal L} = 50~\rm fb^{-1}$ (and $1000~\rm fb^{-1}$) at the LHeC. The signal and background event yields are given by $S = \sigma^{S} \cdot {\cal L}$ and $B = \sigma^{B} \cdot {\cal L}$, where $\sigma^{S}$ and $\sigma^{B}$ are the respective cross sections.

In \autoref{semilepton_cutflow}, for ${\cal L} = 50$ and $1000~{\rm fb}^{-1}$, we show the complete cutflow, beginning with the generator-level selections and progressing through the final set of optimized cuts. The table presents the corresponding event yields for the SM signal and dominant background processes at each stage. The statistical significance, calculated using eq.~\eqref{eq:significance}, is shown in the last row.

In the next \autoref{analysis}, we construct sensitive kinematic observables and estimate bounds on the BSM parameters by performing a $\chi^2$ minimization, using both one-bin and multi-bin analyses of these differential distributions.

\begin{figure*}[t]
    \centering
    \begin{minipage}[b]{0.45\textwidth}
        \includegraphics[width=\textwidth]{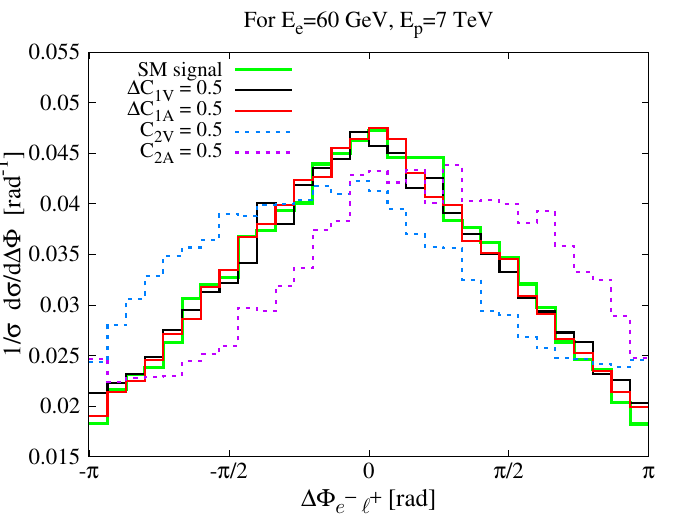}
    \label{delta_phi_e_l}\centerline{(a)}
    \end{minipage}
    \begin{minipage}[b]{0.45\textwidth}
        \includegraphics[width=\textwidth]{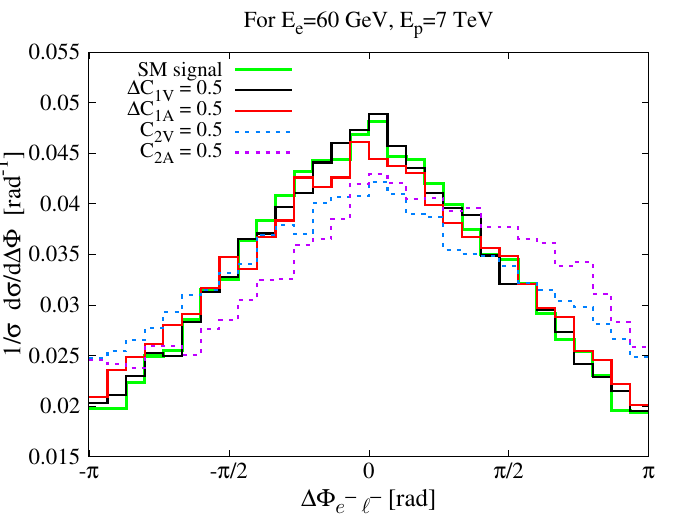}
    \label{delta_phi_e_lminus}\centerline{(b)}
    \end{minipage}
   \hfill
    \begin{minipage}[b]{0.45\textwidth}
        \includegraphics[width=\textwidth]{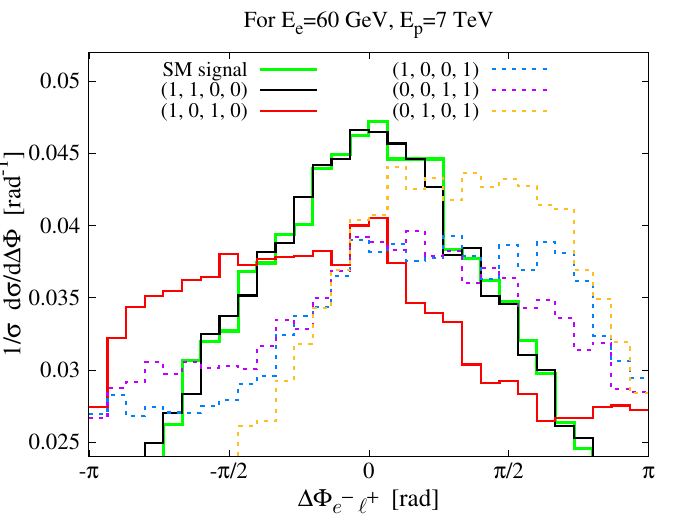}
    \label{delta_phi_e_l}\centerline{(c)}
    \end{minipage}    
    \begin{minipage}[b]{0.45\textwidth}
        \includegraphics[width=\textwidth]{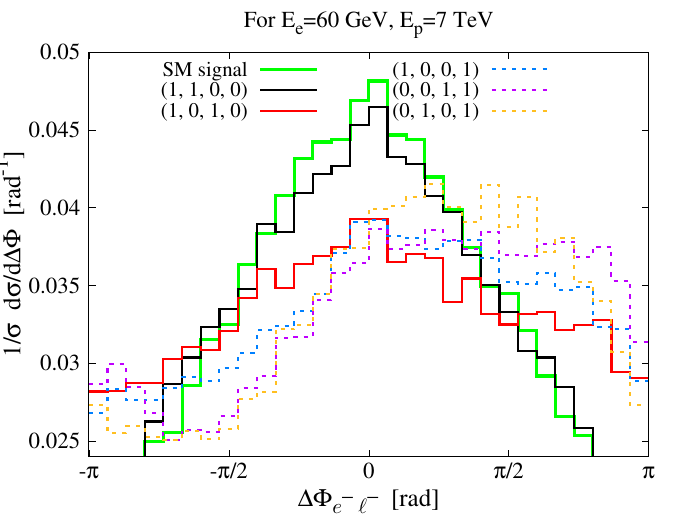}
        \label{delta_phi_two_parameter}\centerline{(d)}
    \end{minipage}
\caption{Normalized differential distributions of the azimuthal angle difference $\Delta \phi_{e^-\,\ell^\pm}$ between the final-state scattered electron $e^{-}$ and the charged lepton $\ell^\pm$, compared with the SM signal. The distributions are shown for benchmark BSM scenarios with coupling values $C_i$: (a) and (b) correspond to varying one coupling at a time ($C_i = 0.5$, others set to zero); (c) and (d) correspond to two couplings set simultaneously to $1.0$, with the remaining set to zero.}
    \label{fig:azimuthal}
\end{figure*}

\begin{figure*}[t]
    \centering
    \begin{minipage}[b]{0.45\linewidth}
        \centering
        \includegraphics[width=\linewidth]{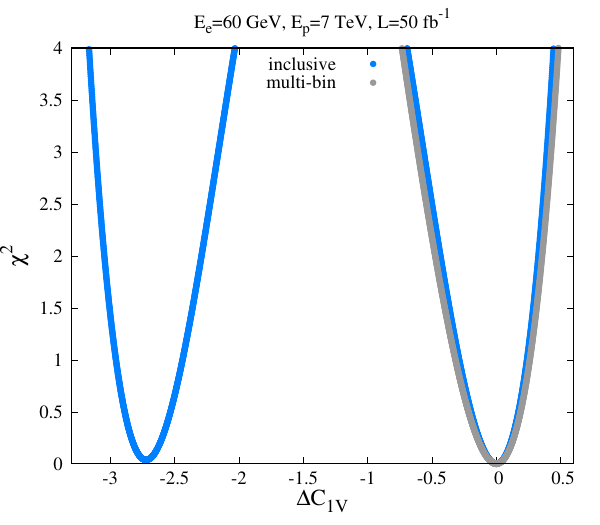}
        \centerline{(a)}
        \label{c1v_bounds}
    \end{minipage}
    \begin{minipage}[b]{0.45\linewidth}
        \centering
        \includegraphics[width=\linewidth]{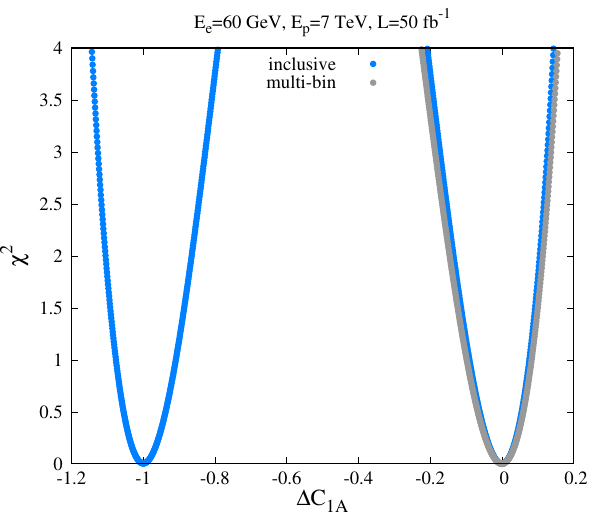}
        \centerline{(b)}
        \label{c1a_bounds}
    \end{minipage}


    \begin{minipage}[b]{0.45\linewidth}
        \centering
        \includegraphics[width=\linewidth]{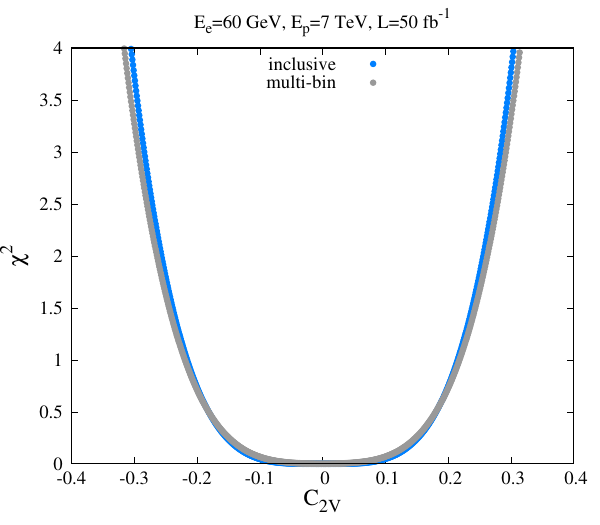}
        \centerline{(c)}
        \label{c2v_bounds}
    \end{minipage}
    \begin{minipage}[b]{0.45\linewidth}
        \centering
        \includegraphics[width=\linewidth]{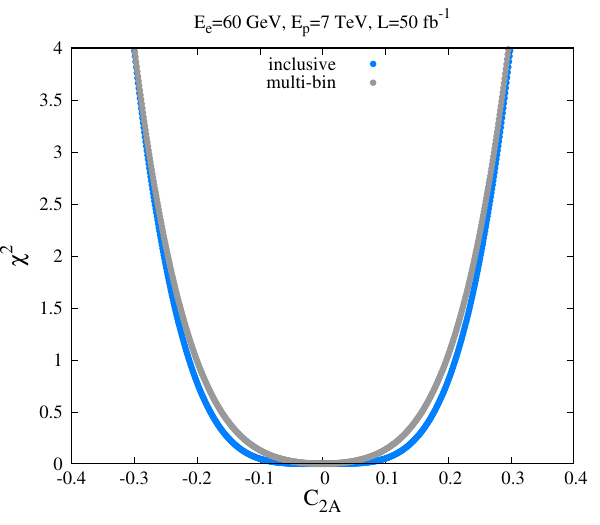}
        \centerline{(d)}
        \label{c2a_bounds}
    \end{minipage}
\caption{The $\chi^2$ distributions obtained from the un-normalized $\Delta \phi_{e^-\,\ell^\pm}$ distribution, used as a sensitive observable for individual $t\bar{t}Z$ coupling parameters. The comparison is shown for the inclusive (web-blue) and differential (web-gray) cases, where the former corresponds to a single-bin (integrated) analysis and the latter to a multi-bin ($n$-bin) analysis. The results are presented for $\sqrt{s} \approx 1.3~\rm TeV$ and $\mathcal{L} = 50~\rm fb^{-1}$, using a threshold of $\chi^2 \approx 4$ to indicate the 95\% confidence level, assuming $\delta_s = 5\%$.}
    \label{one_param_results}
\end{figure*}

\section{Analysis and results}
\label{analysis}

In future $e^-p$ colliders, the initial-state energy asymmetry between the electron and proton beams leads to distinctive angular correlations in the final state~\cite{Biswal:2012mp,Dutta:2013mva,Mosala:2023sse,Coleppa:2017rgb,Ma:2017vve,Rontsch:2014cca,Bouzas:2013jha,Bouzas:2021gwx}. In particular, the difference in azimuthal angle distributions of the scattered electron and one of the final-state leptons or jets originating from the top or anti-top quark decay are sensitive to new physics effects in the $t\bar{t}Z$ couplings. These correlations provide a promising avenue to probe deviations from the SM through kinematic observables constructed from final-state momenta. As shown in \autoref{fig:azimuthal}, the observable $\Delta \phi_{e^- \ell^{\pm}}$ defined as the azimuthal angle difference between the scattered electron and the lepton originating from the top ($\ell^{+}$) or anti-top quark ($\ell^{-}$) decay exhibits notable sensitivity to deviations in the $t \bar{t} Z$ couplings, compared to other possible kinematic pairings. We further exploit this sensitivity by performing a $\chi^2$ analysis on the corresponding differential distributions, considering both one-bin and multi-bin strategies to extract bounds on the BSM parameters. As noted in Ref.~\cite{Mosala:2023sse}, the one-bin (inclusive) $\chi^2$ analysis evaluates the total cross section summed over all bins, thus providing an overall measure of deviations. In contrast, the multi-bin (differential) analysis retains the shape information of the distribution, allowing for a bin-by-bin comparison with the background and yielding improved sensitivity to the BSM couplings through their distinct kinematic behavior. 

The $\chi^2$ is defined as
\begin{equation}
    \chi^2 = \sum_{k=1}^{n} \left(\frac{\sigma_{k}^{\rm BSM}(C_{i}) - \sigma_k^{\rm exp}}{\Delta \sigma_k}\right)^2,
    \label{eq:chi-square1}
\end{equation}
where $\sigma_{k}^{\rm BSM}(C_{i})$ denotes the predicted cross section in the presence of BSM couplings $C_i$, and $\sigma_k^{\rm exp}$ represents the SM expectation in the $k^\text{th}$ bin of a distribution. The total $\chi^2$ is obtained by adding $\chi^2$ of the two channels, i.e., $\chi^2_{\rm tot}=\chi^2_{\ell^-}+\chi^2_{\ell^+}$. In this analysis, we consider only the SM signal as the expected distribution, i.e., $\sigma_k^{\rm exp} = \sigma_k^{\rm SM}$. The uncertainty $\Delta \sigma_{k}$ incorporates both statistical and systematic contributions at a given integrated luminosity $\mathcal{L}$, and is given by
\begin{equation}
    \Delta \sigma_k = \sqrt{\frac{\sigma_k^{\rm SM + Bkg}}{\mathcal{L}} + \left(\delta_{s} \sigma_k^{\rm SM + Bkg}\right)^2}.
    \label{delta_N}
\end{equation}
We assume a systematic uncertainty of $\delta_s = 5\%$ in our analysis, and the $\Delta \phi_{e^- \ell^{\pm}}$ distributions are divided into $n=30$ equally spaced bins. 

The dependence of the inclusive and differential cross sections on the BSM couplings introduced in Lagrangian~\eqref{eq:ttZ-Lag} can be modeled as a quadratic function~\cite{Mosala:2023sse, Rahaman:2019lab, Ravina:2021kpr, Rahaman:2022dwp}:
\begin{equation}
    \sigma^{\rm BSM}(C_i) = \sigma^{\rm SM} + \sum_i C_i A_i + \sum_{i,j} C_i C_j B_{ij},
    \label{eq:cross-section-model}
\end{equation}
where $A_i$ denoting the linear coefficients arising from SM–EFT interference, and $B_{ij}$ representing the quadratic contributions purely from EFT operators. Based on this formalism, we present the results from both one-parameter and two-parameter scans in the following subsection. Note that the limits at ATLAS and CMS presented in \autoref{EFT-theory} are obtained from quadratic fits analogous to eq.~\eqref{eq:cross-section-model}, under the assumption that each parameter is varied independently.

\begin{figure*}[t]
    \centering
    \begin{minipage}[b]{0.496\textwidth}
        \includegraphics[width=\textwidth]{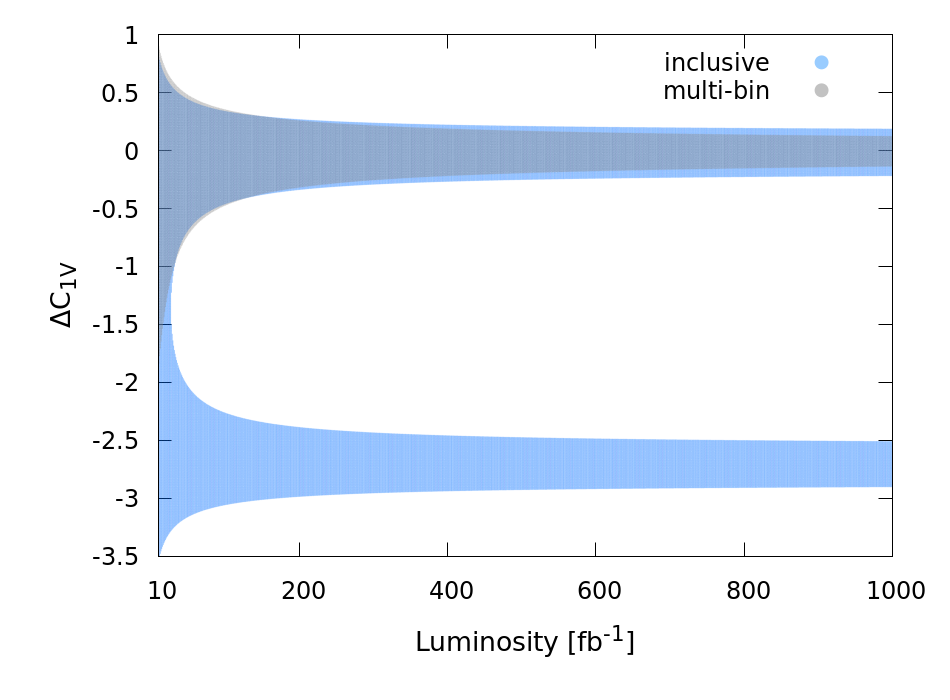}
    \label{sm_luminosity}\centerline{(a)}
    \end{minipage}
    \begin{minipage}[b]{0.496\textwidth}
        \includegraphics[width=\textwidth]{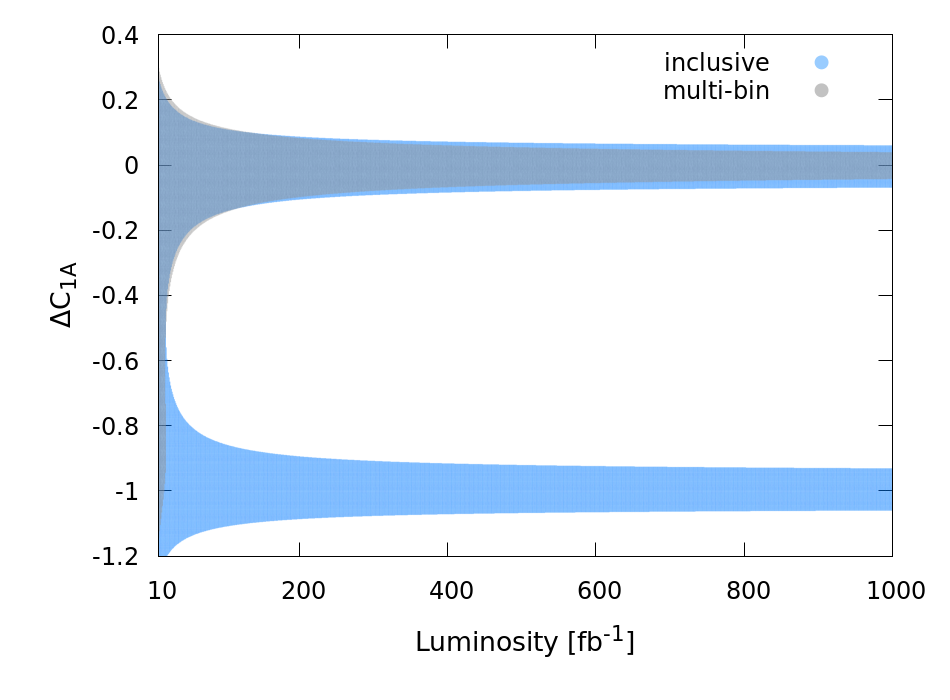}
    \label{bsm_luminosity}\centerline{(b)}
    \end{minipage}
    \hfill
    \begin{minipage}[b]{0.496\textwidth}
        \includegraphics[width=\textwidth]{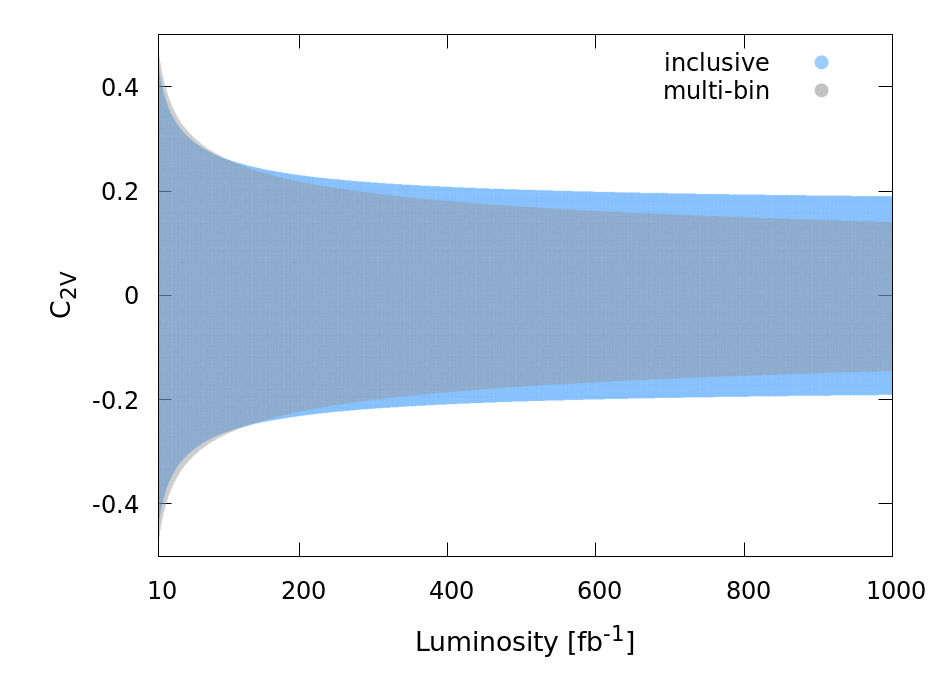}
    \label{sm_luminosity}\centerline{(c)}
    \end{minipage}
   \hfill
    \begin{minipage}[b]{0.496\textwidth}
        \includegraphics[width=\textwidth]{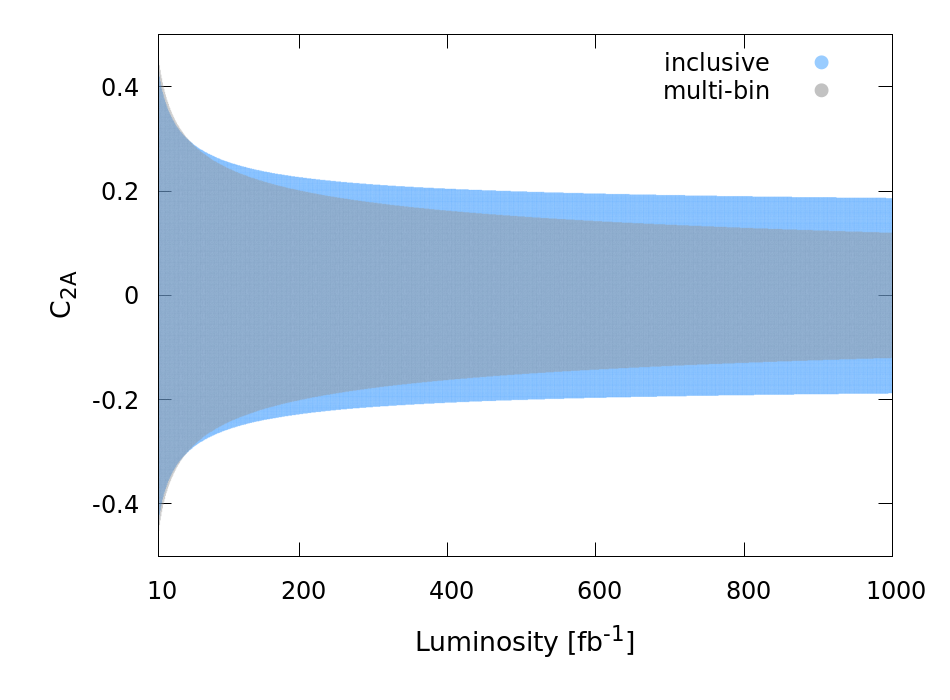}
    \label{bsm_luminosity}\centerline{(d)}
    \end{minipage}
    \caption{Projected 95\% C.L. constraints on the individual BSM couplings as a function of the integrated luminosity, ranging from $10~\mathrm{fb}^{-1}$ to $1000~\mathrm{fb}^{-1}$, obtained from the one-parameter analysis using the inclusive-level (inclusive) and differential-level (multi-bin) approach, assuming $\delta_s = 5\%$. The results illustrate the improvement in sensitivity with increasing luminosity.}
    \label{fig:one_parameter_luminosity}
\end{figure*}

\begin{figure*}[t]
    \centering

    \begin{minipage}{0.6\textwidth}
        \centering
        \includegraphics[width=\linewidth]{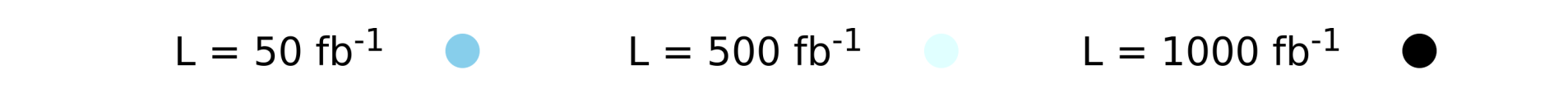}
    \end{minipage}


    \begin{minipage}{0.32\textwidth}
        \centering
        \includegraphics[width=\linewidth]{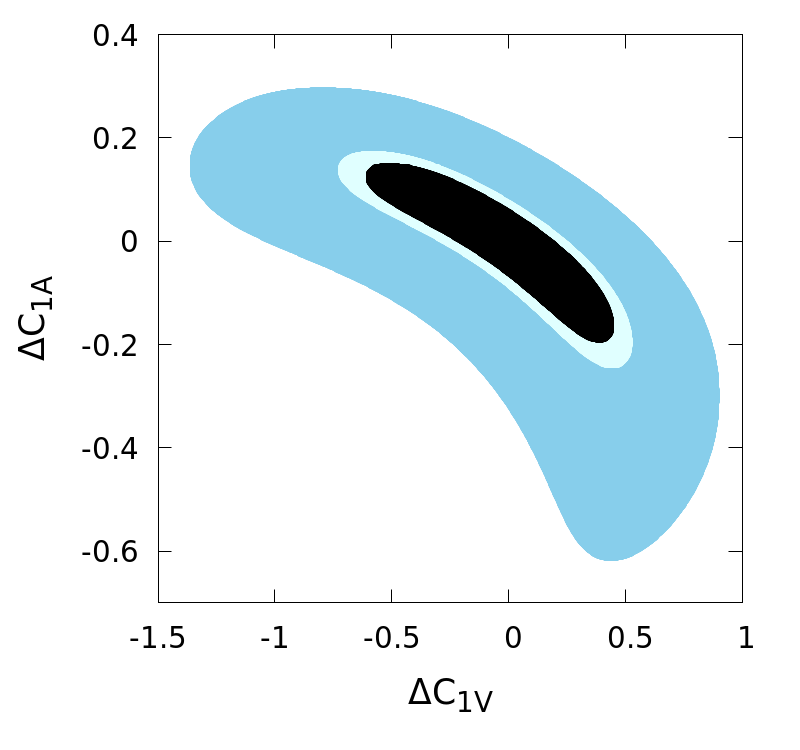}
        \centerline{(a)}
    \end{minipage}
    \hfill
    \begin{minipage}{0.32\textwidth}
        \centering
        \includegraphics[width=\linewidth]{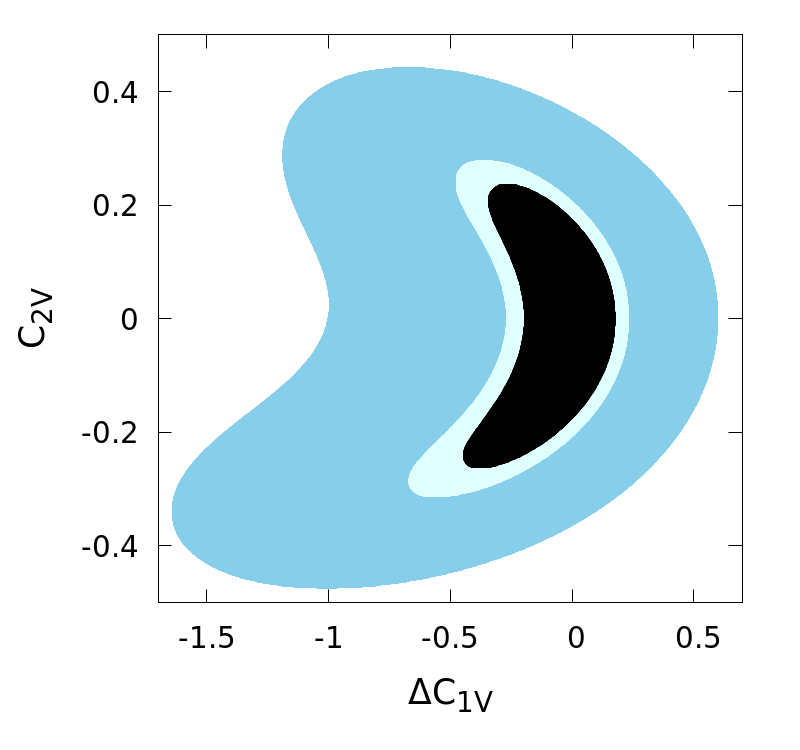}
        \centerline{(b)}
    \end{minipage}
    \hfill
    \begin{minipage}{0.32\textwidth}
        \centering
        \includegraphics[width=\linewidth]{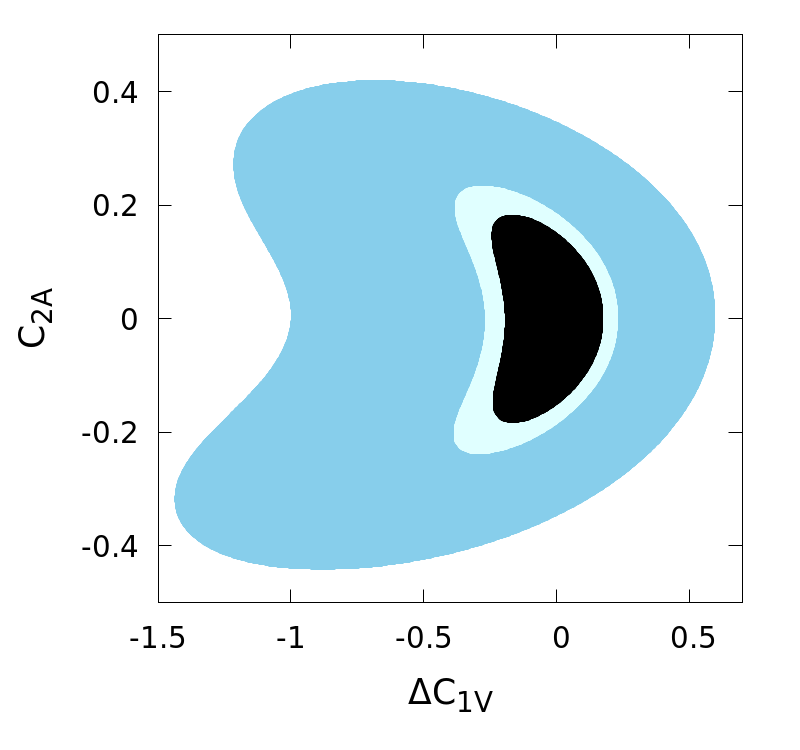}
        \centerline{(c)}
    \end{minipage}

    \vspace{0.5cm}

    \begin{minipage}{0.32\textwidth}
        \centering
        \includegraphics[width=\linewidth]{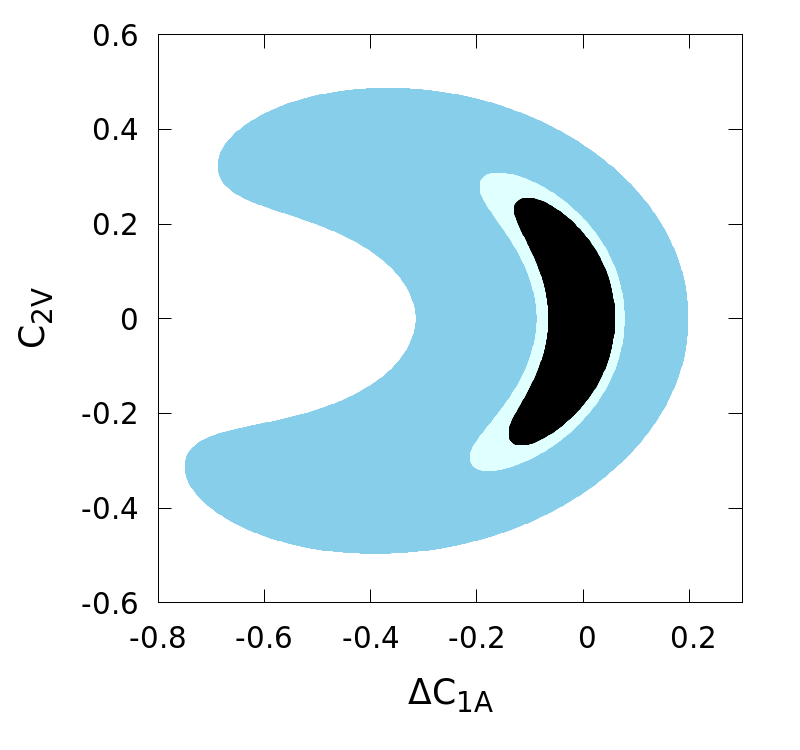}
        \centerline{(d)}
    \end{minipage}
    \hfill
    \begin{minipage}{0.32\textwidth}
        \centering
        \includegraphics[width=\linewidth]{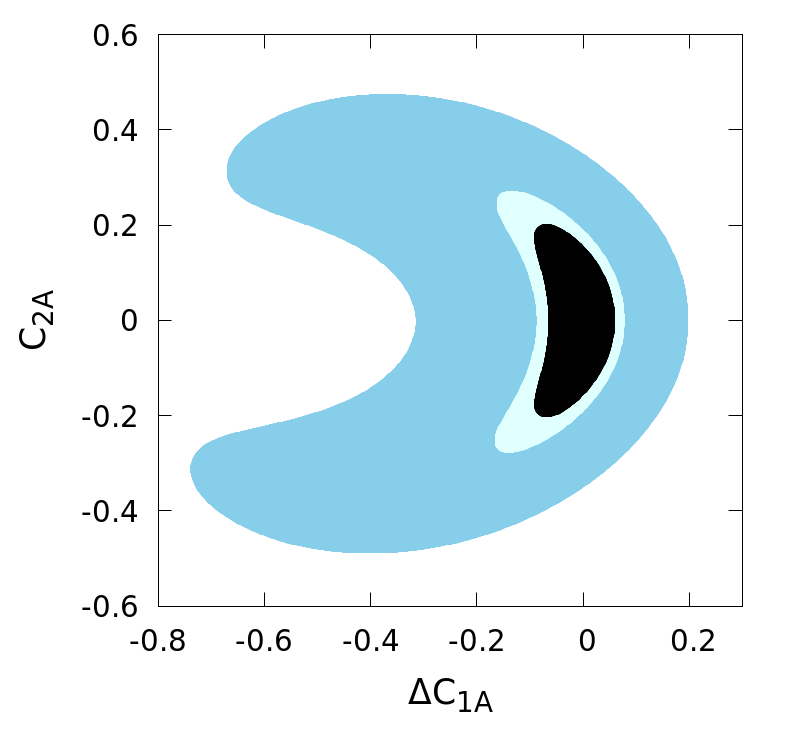}
        \centerline{(e)}
    \end{minipage}
    \hfill
    \begin{minipage}{0.32\textwidth}
        \centering
        \includegraphics[width=\linewidth]{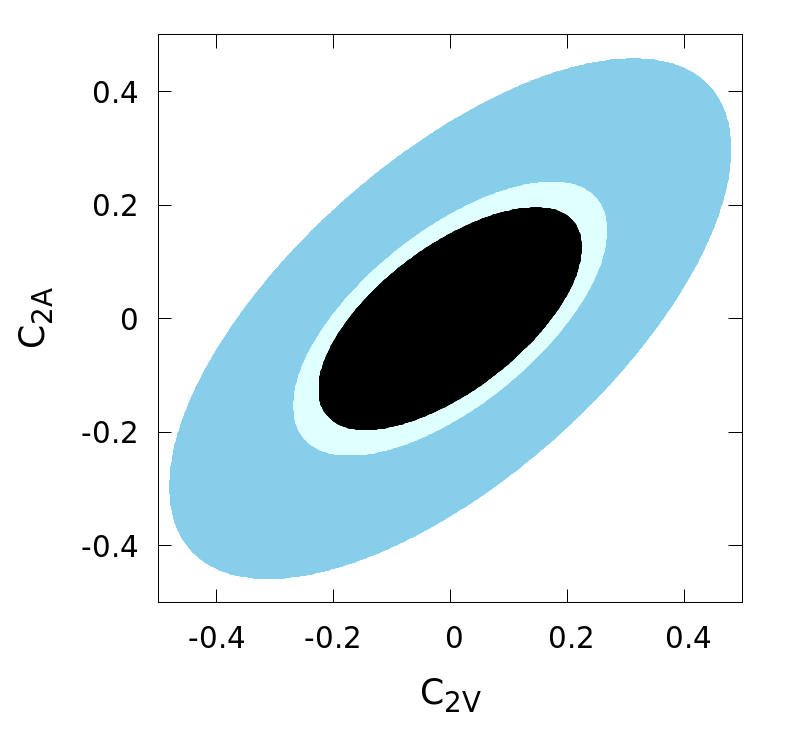}
        \centerline{(f)}
    \end{minipage}

\caption{Allowed parameter space at 95\% C.L. from the two-parameter fit, based on a multi-bin $\chi^2$ analysis in the differential case, which utilizes the full shape of the $\Delta\phi_{e^-\,\ell^\pm}$ distribution across multiple bins, assuming $\delta_s = 5\%$. Results are shown for integrated luminosities of $50~\mathrm{fb}^{-1}$, $500~\mathrm{fb}^{-1}$, and $1000~\mathrm{fb}^{-1}$ at $\sqrt{s} \approx 1.3$~TeV.}
    \label{two_param_luminosity_contours}
\end{figure*}

\subsection{One-parameter analysis\label{one_param}}

To probe the sensitivity of individual BSM couplings at the LHeC, we begin with a one-parameter analysis, where a single coupling $C_i$ is varied and all others are fixed to zero. This provides a baseline estimate of the reach and helps isolate the impact of each parameter on the $t\bar{t}Z$ production process. \autoref{fig:azimuthal} shows the normalized sensitive $\Delta\phi$ distributions for selected benchmark scenarios.  Panels (a) and (b) illustrate cases with $C_{i} = 0.5$, highlighting deviations from the SM induced by individual couplings. 

We then quantify the statistical sensitivity using a $\chi^2$ analysis, defined in eq.\eqref{eq:chi-square1}, with predicted cross sections obtained from eq.\eqref{eq:cross-section-model}. The un-normalized $\Delta \phi$ distribution serves as the sensitive observable. Two analysis strategies are employed: the inclusive case, which uses the integrated cross section (single-bin), and the differential case, which utilizes the full distribution across $n$ bins. The resulting $\chi^2$ distributions are shown in \autoref{one_param_results} for $\sqrt{s} \approx 1.3~\rm TeV$ and $\mathcal{L} = 50~\rm fb^{-1}$, with a threshold of $\chi^2 \approx 4$ corresponding to the 95\% C.L.

In the inclusive analysis, the $\chi^{2}$ distributions exhibit degeneracies with no unique global minima for all BSM parameters. This behavior originates from the quadratic terms $\sum_{i,j} C_i C_j B_{ij}$ in eq.~\eqref{eq:cross-section-model}, which can cause multiple coupling values to produce similar total cross sections, thereby obscuring distinct parameter determination. The interference terms, linear in $C_i$ and represented by $A_i$, are insufficient in the inclusive case to break these degeneracies. These observations are consistent with the features of the total cross section shown in \autoref{inclusive_xsecs}.

By contrast, the differential analysis leverages the full shape information of the un-normalized $\Delta \phi$ distributions, which is more sensitive to subtle changes in the coupling parameters. This additional sensitivity effectively lifts the degeneracies present in the inclusive approach for parameters such as $\Delta C_{1V}$ and $\Delta C_{1A}$, providing well-defined minima in the $\chi^{2}$ profiles. Furthermore, the sensitivity breaks the bifurcated structure in $\Delta C_{1V}$ and $\Delta C_{1A}$ observed in the inclusive approach. This highlights the enhanced discriminating power obtained by utilizing differential observables rather than integrated rates alone. The allowed regions for the BSM couplings show a significant gain in sensitivity when moving from the inclusive to the differential analysis. The parameter space tightens by $46.7\%$ for $\Delta C_{1V}$ and $45.7\%$ for $\Delta C_{1A}$, with the differential analysis eliminating the bifurcation observed for both couplings. The couplings $C_{2V}$ and $C_{2A}$ show negligible change in the allowed parameter space at the luminosity of $50 ~\rm fb^{-1}$ when moving from the inclusive to the differential analysis, mainly due to the limited statistics at lower luminosities, which reduce the sensitivity of the differential distributions and lead to comparable limits in both cases.

\autoref{fig:one_parameter_luminosity} shows the projected constraints on the BSM couplings as a function of the integrated luminosity ranging from $10~\mathrm{fb}^{-1}$ to $1000~\mathrm{fb}^{-1}$, at 95\% confidence level. The allowed parameter space for all BSM couplings decreases with increasing luminosity. In particular, the constraints on $\Delta C_{1V}$ and $\Delta C_{1A}$ exhibit a bifurcated structure for luminosities above $\approx$ 20–25 $~\mathrm{fb}^{-1}$, as shown for the inclusive analysis in panels (a) and (b). Notably, the multi-bin analysis outperforms the inclusive analysis across the entire luminosity range for the vector and axial couplings, whereas the inclusive analysis dominates for $C_{2V}$ and $C_{2A}$ for luminosities below $\approx 100~\rm fb^{-1}$ and $50~\rm fb^{-1}$, respectively, as shown in panels (c) and (d) of \autoref{fig:one_parameter_luminosity}.


Relative to the limits at $50~\mathrm{fb}^{-1}$, the bounds at $1000~\mathrm{fb}^{-1}$ improve by $78.5\%$, $78.9\%$, $54\%$, and $60\%$ for $\Delta C_{1V}$, $\Delta C_{1A}$, $C_{2V}$, and $C_{2A}$, respectively.

\begin{table*}
\caption{The 95\% C.L. limits on the BSM couplings and Wilson coefficients (with $\Lambda = 1~\mathrm{TeV}$), extracted from a multi-bin $\chi^{2}$ analysis of the $\Delta \phi_{e^-\,\ell^\pm}$ distribution in the differential case, which leverages the full shape of the observable across multiple bins. Assuming $\delta_s = 5\%$, results from the one-parameter analysis are shown for $\mathcal{L} = 50~\mathrm{fb}^{-1}$ and $1000~\mathrm{fb}^{-1}$, while the two-parameter analysis includes $\mathcal{L} = 50~\mathrm{fb}^{-1}$, $500~\mathrm{fb}^{-1}$, and $1000~\mathrm{fb}^{-1}$.}
\label{tab:constraints}
\begin{ruledtabular}
\begin{tabular}{lccccc}
& \multicolumn{2}{c}{One-parameter analysis} & \multicolumn{3}{c}{Two-parameter analysis} \\
\cline{2-3} \cline{4-6}
BSM Coupling  &  $50~\mathrm{fb}^{-1}$ & $1000~\mathrm{fb}^{-1}$ &  $50~\mathrm{fb}^{-1}$ & $500~\mathrm{fb}^{-1}$ & $1000~\mathrm{fb}^{-1}$ \\ 
\hline

$\Delta C_{1V}$ & $[-0.733, ~0.482]$ & $[-0.137, ~0.125]$ & $[-1.617, ~0.875]$ & $[-0.702, ~0.502]$ & $[-0.585, ~0.424]$ \\
$\Delta C_{1A}$ & $[-0.224, ~0.155]$ & $[-0.042, ~0.039]$ & $[-0.738, ~0.286]$ & $[-0.233, ~0.163]$ & $[-0.184, ~0.139]$ \\
$C_{2V}$ & $[-0.316, ~0.314]$ & $[-0.145, ~0.141]$ & $[-0.483, ~0.474]$ & $[-0.307, ~0.295]$ & $[-0.253, ~0.242]$ \\
$C_{2A}$ & $[-0.299, ~0.296]$ & $[-0.119, ~0.120]$ & $[-0.475, ~0.461]$ & $[-0.264, ~0.258]$ & $[-0.190, ~0.188]$ \\

\hline
Wilson Coefficient & & & & & \\
\hline

$c_{\phi t} /\Lambda^{2}~(\rm TeV^{-2})$ & $[-5.263, ~7.907]$ & $[-1.355, ~1.479]$ & $[-9.593, ~19.458]$ & $[-5.494, ~7.725]$ & $[-4.652, ~6.354]$ \\
$c_{\phi Q}^{-} /\Lambda^{2}~(\rm TeV^{-2})$ & $[-2.702, ~4.205]$ & $[-0.711, ~0.785]$ & $[-4.866, ~7.263]$ & $[-2.801, ~3.875]$ & $[-2.355, ~3.313]$ \\
$c_{t Z} /\Lambda^{2}~(\rm TeV^{-2})$ & $[-3.692, ~3.669]$ & $[-1.694, ~1.648]$ & $[-5.644, ~5.539]$ & $[-3.587, ~3.447]$ & $[-2.956, ~2.828]$ \\
$c_{t Z}^{I}/\Lambda^{2}~(\rm TeV^{-2})$ & $[-3.494, ~3.459]$ & $[-1.390, ~1.402]$ & $[-5.550, ~5.387]$ & $[-3.085, ~3.015]$ & $[-2.220, ~2.197]$ \\

\end{tabular}
\end{ruledtabular}
\end{table*}

\subsection{Two parameter analysis\label{two_param}}

To capture the effects of parameter correlations and possible interference, we next perform a two-parameter analysis. In this approach, two couplings are varied simultaneously while the remaining couplings are set to zero. This setup provides a more realistic projection by allowing degeneracies and correlations among the couplings to manifest features that typically arise in global fits. Panel (c) and (d) of \autoref{fig:azimuthal} show the normalized $\Delta\phi_{e^-\,\ell^\pm}$  distributions for benchmark scenarios in which two couplings are simultaneously set to $1.0$. These illustrate the potential distortions introduced by correlated BSM effects. As observed in the one-parameter analysis, the multi-bin approach yields stronger constraints on the BSM couplings. Therefore, we adopt and present results based exclusively on the multi-bin analysis in what follows. 

Using the linear and quadratic coefficients extracted from the one-parameter fits of eq.~\eqref{eq:cross-section-model}, we determine the off-diagonal EFT quadratic coefficients $B_{ij}$ for $i \neq j$. The allowed parameter space from the two-parameter analysis is shown in \autoref{two_param_luminosity_contours}, based on the multi-bin $\chi^2$ method applied to the unnormalized $\Delta\phi$ distributions. Representative results are presented for integrated luminosities of $50~\mathrm{fb}^{-1}$, $500~\mathrm{fb}^{-1}$, and $1000~\mathrm{fb}^{-1}$, with the confidence regions defined by $\chi^2 \approx 6$, corresponding to the 95\% C.L. 

We observe that $\Delta C_{1V}$ exhibits a strong negative correlation with $\Delta C_{1A}$, as seen in panel (a) of \autoref{two_param_luminosity_contours}, and a weak correlation with $C_{2V}$ and $C_{2A}$ (panels (b) and (c)). Similarly, $\Delta C_{1A}$ is weakly correlated to the anomalous weak magnetic and electric dipole couplings across all luminosities, as seen in panels (d) and (e). In contrast, $C_{2V}$ and $C_{2A}$ (panel~(f)) shows strong correlations across all luminosity scenarios. As expected, the contours shrink with increasing luminosity, reflecting improved statistical sensitivity.

Compared to the one-parameter analysis, the two-parameter fit yields significantly broader allowed regions due to correlations between parameters. Considering the bounds for $\Delta C_{1V}$ and $\Delta C_{1A}$ from the two-parameter analysis as shown in \autoref{two_param_luminosity_contours}~(a), (b) and (d), and for $C_{2V}$ and $C_{2A}$ from panels (d) and (e), respectively, the allowed parameter space at $50~\mathrm{fb}^{-1}$ ($1000~\mathrm{fb}^{-1}$) enlarges by 105.8\% (288.5\%),  168.4\% (300.0\%), 52.4\% (72.4\%), and 56.7\% (58.3\%) for $\Delta C_{1V}$, $\Delta C_{1A}$, $C_{2V}$, and $C_{2A}$, respectively, when transitioning from the one-parameter to two-parameter analysis.
The corresponding constraints on the BSM couplings and WCs, with $\Lambda = 1~\mathrm{TeV}$ (using eq.~\eqref{eq:cwc_all}), are summarized in \autoref{tab:constraints} for both the one- and two-parameter analyses across the selected luminosities.

To further quantify the interplay among the parameters, we estimate the statistical correlations between the BSM couplings. We adopt the Hessian method~\cite{Pumplin:2000vx, Pumplin:2001ct}, computing the Hessian matrix, $H_{ij}$, of the $\chi^{2}$ function in eq.~\eqref{eq:chi-square1}, evaluated at a point $C^{\rm fit}$ near the global minimum (best fit). The Hessian is defined as~\cite{Easther:2016ire}:
\begin{equation}
    H_{ij} = \frac{\partial^{2} \chi^{2}}{\partial C_{i}\partial C_{j}} \Bigg | _{\bar{C} = C^{\rm fit}}.
\end{equation}

The covariance matrix, $\mathcal{C}_{ij}$, is then obtained as the inverse of the Hessian, i.e., $\mathcal{C}_{ij}  = [H^{-1}]_{ij}$. From this, the correlation matrix is extracted as:
\begin{equation}
    \rho_{ij} = \frac{\mathcal{C}_{ij}}{\sqrt{\mathcal{C}_{ii} \mathcal{C}_{jj}}}.
\end{equation}

The correlation matrix shown in \autoref{fig:correlation} summarizes the interplay between the couplings. It is evident that a strong inverse correlation exists between $ \Delta C_{1V} $ and $\Delta C_{1A}$, suggesting that constraining one will aid in constraining the other. A slight correlation of $C_{2A}$ with both $\Delta C_{1V}$ ($-10\%$) and $\Delta C_{1A}$ ($10\%$) indicate minimal interplay in their sensitivities. Additionally, $C_{2V}$ shows minimal correlations with $\Delta C_{1V}$ ($-9.9$\%) and $\Delta C_{1A}$ ($9.8\%$), while it remains strongly correlated with $\Delta C_{2A}$ ($77\%$). These correlations reflect the extent to which the uncertainty in each parameter is influenced by the others in the multidimensional parameter space.

In both the one- and two-parameter analyses, we have adopted a conservative approach by varying individual or pairs of BSM coupling parameters at a time, while keeping the remaining coefficients fixed, assuming they are already tightly constrained by existing data. This strategy enables a clearer interpretation of the effects of specific parameters. However, for more realistic projections, a comprehensive global analysis involving multiple non-zero parameters is essential. In the following subsection, we present the results of a multi-parameter simultaneous fit, where all couplings are allowed to vary simultaneously.
\begin{figure}
    \centering
    \includegraphics[width=0.9\linewidth]{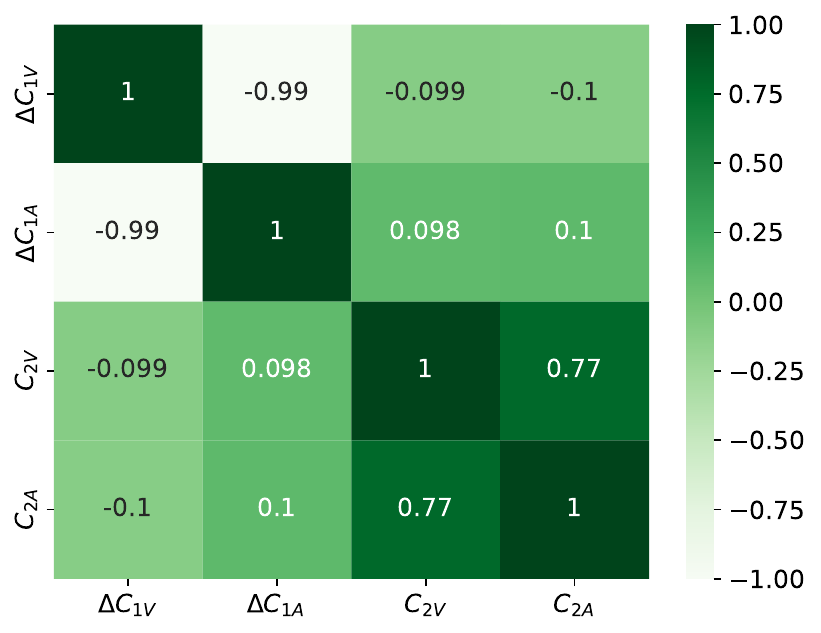}
    \caption{Correlations between BSM couplings in the two-parameter case, extracted from the multi-bin $\chi^{2}$ analysis. Results are shown for $\delta_s = 5\%$ at $\sqrt{s} \approx 1.3~\mathrm{TeV}$.}
    \label{fig:correlation}
\end{figure}

\subsection{Multi-parameter simultaneous analysis}
\label{sec:mcmc_analysis}

\begin{figure}[t!]
\centering
\includegraphics[width=1\linewidth]{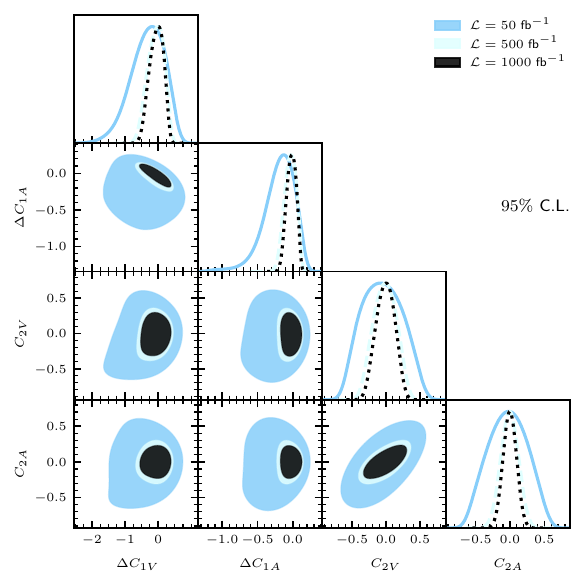}
\caption{Marginalised one-dimensional and two-dimensional projections at 95\% C.L. from the MCMC analysis are shown in a triangular array for the anomalous couplings, using the binned $\Delta\phi_{e^-\ell^\pm}$ distribution with $\delta_s = 5\%$, for integrated luminosities of $50~\mathrm{fb}^{-1}$, $500~\mathrm{fb}^{-1}$, and $1000~\mathrm{fb}^{-1}$ at $\sqrt{s} \approx 1.3$~TeV.}
\label{fig:mcmc95}
\end{figure}

In order to obtain robust constraints on the BSM couplings, we perform a multi-parameter analysis by simultaneously varying all parameters and extracting the posterior distributions after marginalization using a Markov Chain Monte Carlo (MCMC) approach. The analysis is based on the binned $\Delta\phi_{e^- \ell^\pm}$ distributions, incorporating a systematic uncertainty of $\delta_s = 5\%$. We employ the \texttt{GetDist} package~\cite{Lewis:2019xzd} to carry out the marginalization and to obtain the posterior projections and credible intervals for the BSM couplings. This approach allows for a more realistic estimation of parameter constraints by properly accounting for correlations and degeneracies among the couplings in the multi-dimensional parameter space.

Figure~\ref{fig:mcmc95} shows the one- and two-dimensional marginalized posterior distributions for the anomalous couplings $\Delta C_{1V}$, $\Delta C_{1A}$, $C_{2V}$, and $C_{2A}$. The shaded regions indicate the 95\% C.L. constraints, shown for three different integrated luminosities: $50~\mathrm{fb}^{-1}$, $500~\mathrm{fb}^{-1}$, and $1000~\mathrm{fb}^{-1}$. The corresponding simultaneous limits on the couplings are summarized in Table~\ref{tab:mcmc95}.
\begin{table}
\caption{The 95\% C.L. limits on the BSM couplings and corresponding Wilson coefficients (with $\Lambda = 1~\mathrm{TeV}$), obtained from the MCMC analysis using the binned $\Delta \phi_{e^-\,\ell^\pm}$ distribution. Simultaneous constraints are shown for $\delta_s = 5\%$ and integrated luminosities of $\mathcal{L} = 50~\mathrm{fb}^{-1}$, $500~\mathrm{fb}^{-1}$, and $1000~\mathrm{fb}^{-1}$.}
\label{tab:mcmc95}
\centering
\begin{ruledtabular}
\begin{tabular}{lccc}
Coupling& $50~\mathrm{fb}^{-1}$ & $500~\mathrm{fb}^{-1}$ & $1000~\mathrm{fb}^{-1}$ \\ \hline
 $\Delta C_{1V} $ & $[-1.318, ~0.592]$ & $[-0.543, ~0.372]$ & $[-0.443, ~0.318]$ \\
 $\Delta C_{1A} $ & $[-0.625, ~0.187]$ & $[-0.195, ~0.116]$ & $[-0.148, ~0.102]$ \\
 $C_{2V}        $ & $[-0.567, ~0.486]$ & $[-0.330, ~0.292]$ & $[-0.269, ~0.242]$ \\
 $C_{2A}        $ & $[-0.552, ~0.488]$ & $[-0.253, ~0.243]$ & $[-0.190, ~0.186]$ \\

\hline
WC & & & \\
\hline
$c_{\phi t} /\Lambda^{2}$ & $[-6.436, ~16.054]$ & $[-4.032, ~6.098]$ & $[-3.470, ~4.883]$ \\
$c_{\phi Q}^{-} /\Lambda^{2}$ & $[-3.346, ~5.726]$ & $[-2.115, ~2.875]$ & $[-1.785, ~2.437]$ \\
$c_{t Z} /\Lambda^{2}$ & $[-6.625, ~5.679]$ & $[-3.856, ~3.412]$ & $[-3.143, ~2.828]$ \\
$c_{t Z}^{I} /\Lambda^{2}$ & $[-6.450, ~5.702]$ & $[-2.956, ~2.839]$ & $[-2.220, ~2.173]$ \\
\end{tabular}
\end{ruledtabular}
\end{table}

The diagonal panels of \autoref{fig:mcmc95} display the marginalized one-dimensional posterior distributions, which become progressively narrower with increasing luminosity, reflecting improved sensitivity. In particular, the constraints on $\Delta C_{1V}$ and $\Delta C_{1A}$ become significantly tighter at $1000~\mathrm{fb}^{-1}$, indicating strong sensitivity to these couplings.

The off-diagonal panels show the two-dimensional credible regions for each pair of couplings. Compared to the two-parameter $\chi^2$ analysis in \autoref{two_param_luminosity_contours}, the MCMC approach provides a more comprehensive exploration of the parameter space by fully marginalizing over all other couplings. As a result, the credible contours obtained from MCMC are generally broader than those from the two-parameter projections, which assume all other couplings are fixed to their SM values. While the correlations observed in the MCMC results are somewhat weaker at 50 fb$^{-1}$ of luminosity, they remain qualitatively consistent  for the higher luminosities with those seen in the two-parameter analysis.

Quantitatively, the multi-parameter analysis yields broader confidence intervals for the couplings $C_{2V}$ and $C_{2A}$ compared to the two-parameter results, as expected when correlations among all parameters are taken into account. In contrast, the simultaneous limits obtained for $\Delta C_{1V}$ and $\Delta C_{1A}$ are slightly tighter than those of the two-parameter analysis. These differences arise because, in the multi-parameter fit, the confidence interval for each coupling is obtained after marginalizing over all others, whereas in the two-parameter case, both parameters are allowed to vary simultaneously, and the limits correspond to their maximum allowed ranges without marginalization.

\section{Summary and Conclusions}
\label{sec:Conc}

In this study, we have investigated the sensitivity of future $e^- p$ colliders, such as the LHeC, to both precision deviations in the SM $t\bar{t}Z$ couplings ($\Delta C_{1V}$, $\Delta C_{1A}$) and potential anomalous contributions from physics beyond the SM ($C_{2V}$, $C_{2A}$). We consider electron and proton beam energies of $E_e = 60~\mathrm{GeV}$ and $E_p = 7~\mathrm{TeV}$, corresponding to a center-of-mass energy of $\sqrt{s} \approx 1.3~\mathrm{TeV}$, with the electron beam polarized at $-80\%$. Our analysis focused on the neutral current process $e^- p \to e^- t \bar{t}$, considering the semileptonic final state, which offers a clean experimental signature with manageable backgrounds. We derived 95\% confidence level bounds on the anomalous $t\bar{t}Z$ coupling parameters using both inclusive and differential cross-section analyses, for both the one-parameter and two-parameter scenarios. The results presented assume a fixed systematic uncertainty of $\delta_s = 5\%$.

We demonstrate the potential of the azimuthal angle distribution between the scattered electron and one of the final-state leptons originating from the top-quark decays ($\Delta \phi_{e^{-} \ell^{\pm}}$), to enhance sensitivity beyond what is achievable with the total cross section alone. 

The differential analysis improves sensitivity over the inclusive case, tightening the allowed regions by approximately 46.7\% for $\Delta C_{1V}$, and 45.7\% for $\Delta C_{1A}$. As the integrated luminosity increases from $50~\mathrm{fb}^{-1}$ to $1000~\mathrm{fb}^{-1}$, the bounds on $\Delta C_{1V}$ and $\Delta C_{1A}$ improve by approximately 78.5\% and 78.9\%, whereas those on $C_{2V}$ and $C_{2A}$ improve by 54\% and 60\%, respectively. The two-parameter analysis results in broader allowed regions due to parameter correlations, yielding constraints more than twice as loose compared to the one-parameter case. In contrast, the multi-parameter analysis, which accounts for simultaneous variation and marginalization over all couplings, produces constraints that are generally broader than the two-parameter results. This reflects the full impact of parameter correlations and degeneracies in the multidimensional space, providing a more conservative but realistic estimate of the allowed parameter regions.
  
\begin{figure}[t]
    \centering
    \includegraphics[width=1\linewidth]{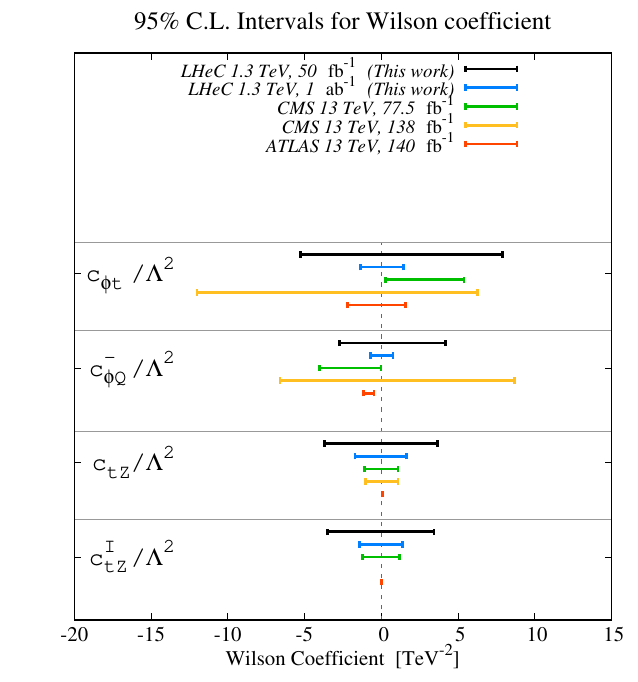}
    \caption{The $95\%$ C.L. intervals for the Wilson coefficients, with $\Lambda = 1~\mathrm{TeV}$, extracted from a multi-bin $\chi^{2}$ analysis of the $\Delta \phi_{e^-\,\ell^\pm}$ distribution in the differential case, which leverages the full shape of the observable across multiple bins. Results are shown for the one-parameter analysis at $50~\mathrm{fb}^{-1}$ (black) and $1000~\mathrm{fb}^{-1}$ (blue) for $\sqrt{s} \approx 1.3~\mathrm{TeV}$, assuming $\delta_s = 5\%$ ({\it This work}). For comparison, observed limits from CMS at $77.5~\mathrm{fb}^{-1}$ (green) and $138~\mathrm{fb}^{-1}$ (yellow), and from ATLAS at $140~\mathrm{fb}^{-1}$ (red), are shown for $\sqrt{s} = 13~\mathrm{TeV}$.}
\label{wilson_comparison}
\end{figure}

Translating the bounds on $\Delta C_{1V}$ and $\Delta C_{1A}$ into absolute values of the couplings, defined as $C_{1V} \equiv C_{1V}^{\rm SM} + \Delta C_{1V}$ and $C_{1A} \equiv C_{1A}^{\rm SM} + \Delta C_{1A}$, we find that at $1000~\mathrm{fb}^{-1}$ (95\% C.L), the couplings are constrained to:
\begin{align}
    C_{1V} = 0.192^{+0.125}_{-0.137},  \quad
    C_{1A} = 0.500^{+0.039}_{-0.042}, \notag
\end{align}
corresponding to projected average relative precisions of approximately 68\% and 8\%, respectively, with respect to their SM values, based on the one-parameter analysis.

We have also obtained projected constraints on the Wilson coefficients relevant to the $t\bar{t}Z$ couplings considered in this study. Compared to our results, the earlier LHeC projections reported in Ref.~\cite{Bouzas:2013jha}, which are based solely on the total cross section as the observable, yield significantly weaker bounds at twice the luminosity ($100~\rm fb^{-1}$) we have considered. On the hand, our projected constraints on $c_{\phi t}$, $c_{\phi Q}^{-}$ at $1000~\rm fb^{-1}$ are significantly better than future projection at HL-LHC~\cite{MammenAbraham:2022yxp} (taken up to the quadratic term), while $c_{t Z}$ and $c_{t Z}^{I}$ are comparable. The ILC~\cite{Rontsch:2015una} constraints indicate stronger sensitivity to all WCs.

A comparative analysis of the projected 95\% C.L. limits on the WCs at the LHeC (this work), based on the one-parameter analysis, and existing constraints from the LHC experiments (ATLAS and CMS) is presented in \autoref{wilson_comparison}. At an integrated luminosity of $50~\mathrm{fb}^{-1}$, the projected bounds on $c_{t Z}$ and $c_{t Z}^{I}$ at the LHeC are generally weaker than current LHC results, which benefit from higher statistics and global fits involving multiple channels.

Interestingly, we find that the CMS constraints on $c_{\phi t}$ and $c_{\phi Q}^{-}$ at $138~\mathrm{fb}^{-1}$ are approximately 1.4 and 2.2 times weaker than our projections, respectively, while the bounds from CMS at $77.5~\mathrm{fb}^{-1}$ are generally stronger than ours. At the high-luminosity phase of the LHeC with $1000~\mathrm{fb}^{-1}$, the sensitivity is expected to improve substantially, yielding stronger constraints on $c_{\phi t}$ and $c_{\phi Q}^{-}$ than those from CMS, and comparable constraints on $c_{tZ}$ and $c_{tZ}^{I}$ . However, the limits on $c_{\phi Q}^{-}$ and $c_{tZ}$ ($c_{tZ}^{I}$) remain slightly weaker than those from ATLAS. These results reflect both the enhanced kinematic reach and clean final states of the $e^-p$ environment, as well as the differing sensitivities of various observables to specific couplings.

\bibliography{main}
\bibliographystyle{apsrev4-2}

\end{document}